\documentclass[noshowpacs,amsmath,
twocolumn,
superscriptaddress,
8pt,
aps,prb]{revtex4-1}
\usepackage{setspace}
\usepackage{amsmath}
\usepackage{graphicx}
\usepackage[nearskip,margin = 0pt]{subfig}

\usepackage{verbatim}
\usepackage{amsfonts}
\usepackage{amssymb}
\usepackage{epstopdf} 
\usepackage{xcolor}
\usepackage{verbatim}
\usepackage{upgreek}
\usepackage{ragged2e}
\usepackage{url}
\usepackage[hidelinks]{hyperref}
\DeclareGraphicsExtensions{.pdf,.eps,.png,.jpg,.mps}

\begin{document}
\title{Universal loss and gain characterization inside photonic integrated circuits}


\author{Haoran Chen$^{1,*}$, Ruxuan Liu$^{1,*}$, Gedalia Y. Koehler$^{1,*}$, Fatemehsadat Tabatabaei$^{1}$, Xiangwen Guo$^{1}$, Shuman Sun$^{1}$, Zijiao Yang$^{1,2}$, Beichen Wang$^{1}$, Andreas Beling$^{1}$ and Xu Yi$^{1,2,\dagger}$\\
\vspace{3pt}
$^1$Department of Electrical and Computer Engineering, University of Virginia, Charlottesville, Virginia 22904, USA.\\
$^2$Department of Physics, University of Virginia, Charlottesville, Virginia 22904, USA.\\
$^{*}$These authors contributed equally.\\
$^{\dagger}$Corresponding authors: yi@virginia.edu}

\begin{abstract}
Integrated photonics has undergone tremendous development in the past few decades, transforming many fields of study in science and technology. Loss and gain are two fundamental elements in photonic circuits and have direct impacts on nearly all key performance metrics. Surprisingly, the tools to characterize the optical loss and gain inside photonic integrated circuits (PICs) are very limited. This is because, unlike free-space or fiber optics, integrated circuits cannot be nondestructively disassembled. Here, we report a universal method to see inside the photonic integrated circuits and measure loss and gain on the component level nondestructively. The method leverages nonlinear optical devices as optical power discriminators to retrieve the loss and gain information inside the PICs. 
Our method has a precision better than 0.1 dB, and can characterize the loss of individual fiber-chip coupling facet and general unknown devices under test.
As a demonstration of applications, we measured the true on-chip quantum efficiency of a quantum PIC consisting of heterogeneously integrated balanced photodiodes, a critical building block for integrated quantum technology. Our method can be implemented on different photonic platforms, and can be used to understand gain and loss in complex photonic circuits, which is essential to optimize circuit design and to create large-scale systems with predictable, reproducible performance.

\end{abstract}
\date{\today}

\maketitle

\noindent {\bf Introduction}









Optical loss and gain are of fundamental importance in photonic integrated circuits (PICs) and play critical roles in many scientific fields of study. Gain and loss directly impact the signal-to-noise ratio in optical communications\cite{}, create PT-symmetry phenomena in photonic circuits\cite{ruter2010observation,el2018non}, and affect a wide range of nonlinear optical behaviors \cite{boyd2008nonlinear}. In quantum photonics, optical loss serves as a primary source of decoherence \cite{}, as it degrades quantum entanglement and squeezing\cite{vahlbruch2016detection,andersen201630}, which are critical resources for photonic quantum computing\cite{andersen201630,takeda2019toward,wang2020integrated,zhong2020quantum}, quantum information \cite{pirandola2015advances} and quantum enhanced sensing \cite{tse2019quantum,casacio2021quantum}. In recent years, PICs have experienced steady growth in both physical footprint and component density to support increasingly sophisticated functionality \cite{xiang2021perspective}. Characterizing gain and loss inside large-scale PICs is vital for informed circuit design and optimal circuit performance. 

However, there is no good method to look inside the circuits and accurately identify loss and gain on a component level because photonic integrated circuits cannot be nondestructively disassembled. Typically, only the total loss/gain of a circuit, e.g., fiber-to-fiber loss, can be directly measured. This is true even for circuits with only one photonic component, as loss always exists when light couples in and out of the chip, and the coupling loss varies from coupler to coupler. An exception is the optical backscattering reflectometry (OBR) \cite{barfuss1989modified,soller2005high}, which uses back-reflection in fibers and waveguides to infer optical loss in the circuit. However, it measures loss indirectly and thus has many limitations. So far, its success in PICs has been limited to measuring the loss and dispersion of long waveguides \cite{bauters2011ultra,bauters2011planar,lee2012ultra}.

\medskip

Here, we report a universal method to see inside the photonic integrated circuits and measure loss and gain on the component level nondestructively. Our method uses nonlinear optical devices as optical power discriminators to identify the optical loss/gain difference on the left side and right side of the nonlinear device. The general concept is that if light experiences different losses when it arrives at a nonlinear device from the left path and the right path, then this difference can be reflected by optical nonlinearity phenomena, e.g., optical parametric oscillation (OPO) threshold\cite{kippenberg2004kerr}. The concept is analogous to breaking optical reciprocity through optical nonlinearity and asymmetric optical paths. This additional information, combined with the measurement of total loss/gain of the optical path ($\alpha$ dB), can yield the loss/gain information of each component in the circuit. Our method is nondestructive as we only measure the in-fiber optical power going into the chip and out of the chip, and the added nonlinear device, such as a ring resonator, can be frequency detuned when not in use.

\begin{figure*}[!bht]
\captionsetup{singlelinecheck=off, justification = RaggedRight}
\includegraphics[width=17cm]{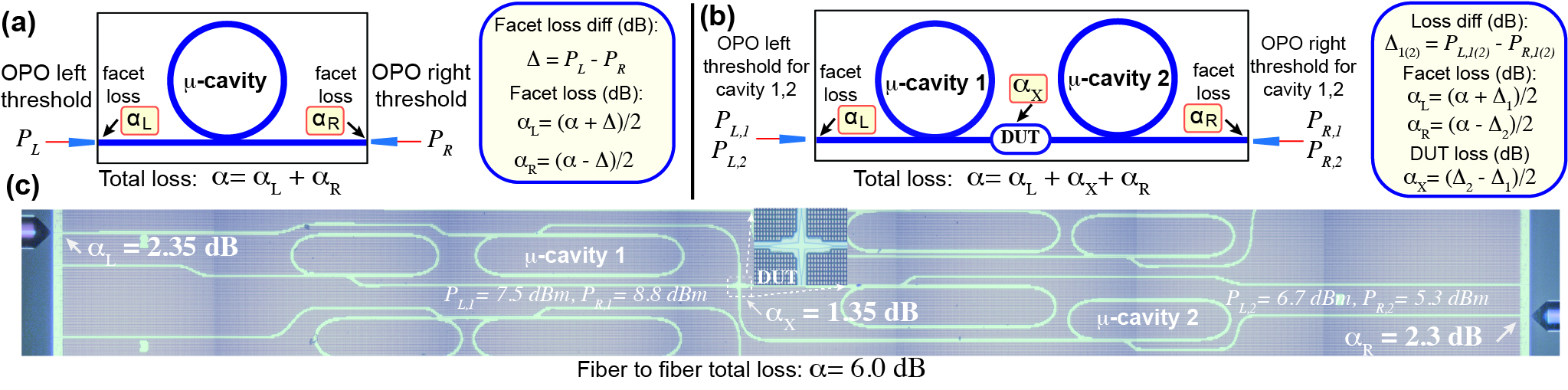}
\caption{{\bf Illustration of universal loss and gain measurement method for photonic integrated circuits.} The method is based on the general concept that if light experiences different losses when it arrives at a nonlinear device from the left path and the right path, then this difference should be reflected by the optical nonlinearity phenomena, e.g., OPO threshold. If $\Delta$ dB more laser power is needed to reach the same nonlinear phenomenon when the laser is injected from the left than from the right, then it means there is $\Delta$ dB more loss on the left-side optical path than on the right-side optical path. The method is illustrated for two general scenarios: 
\textbf{(a)} Fiber chip coupling loss characterization. Measurement of the total optical loss ($\alpha$), and the in-fiber OPO threshold pumping from the left facet ($P_L$) and right facet ($P_R$) can unveil the individual facet loss of the left ($\alpha_L$) and the right ($\alpha_R$). \textbf{(b)} The general method of measuring an unknown loss/gain ($\alpha_X$) of an on-chip device under test (DUT). \textbf{(c)} An example of characterizing individual facet loss and an unknown component loss (waveguide crossing) in the PIC. All variables are in log scale (dBm or dB).}
\label{fig:concept}
\end{figure*}

\medskip

The basic principle of our method works as follows: for a spatially symmetric device, e.g., a ring resonator, the strength of a nonlinear optical phenomenon depends on the laser power going into the device, but not the direction of the laser. As a result, if $\Delta$ dB more laser power is needed to reach the same nonlinear phenomenon when the laser is injected from the left than from the right, then it means there is $\Delta$ dB more loss on the left-side optical path than the right-side optical path. Two general scenarios are used to illustrate this concept: (1) measurements of individual fiber-chip facet coupling loss (Fig.\ref{fig:concept}a) and (2) a general unknown loss measurement for a device under test (DUT) on a chip (Fig.\ref{fig:concept}b). An example to illustrate the power of the method is shown in Fig.\ref{fig:concept}c, where the loss of left coupling facet, device under test (a waveguide crossing), and the right facet loss in the PIC are all identified. The method can be extended to any PICs using the same principle. In this paper, two common nonlinear phenomena in ring resonators are used for loss characterization: Kerr OPO \cite{kippenberg2004kerr}, and resonance frequency drift from thermal-optics effect \cite{carmon2004dynamical}. Other optical nonlinearity shall work as well.

\medskip

\noindent {\bf Results.}

\begin{figure*}[!bht]
\captionsetup{singlelinecheck=off, justification = RaggedRight}
\includegraphics[width=17cm]{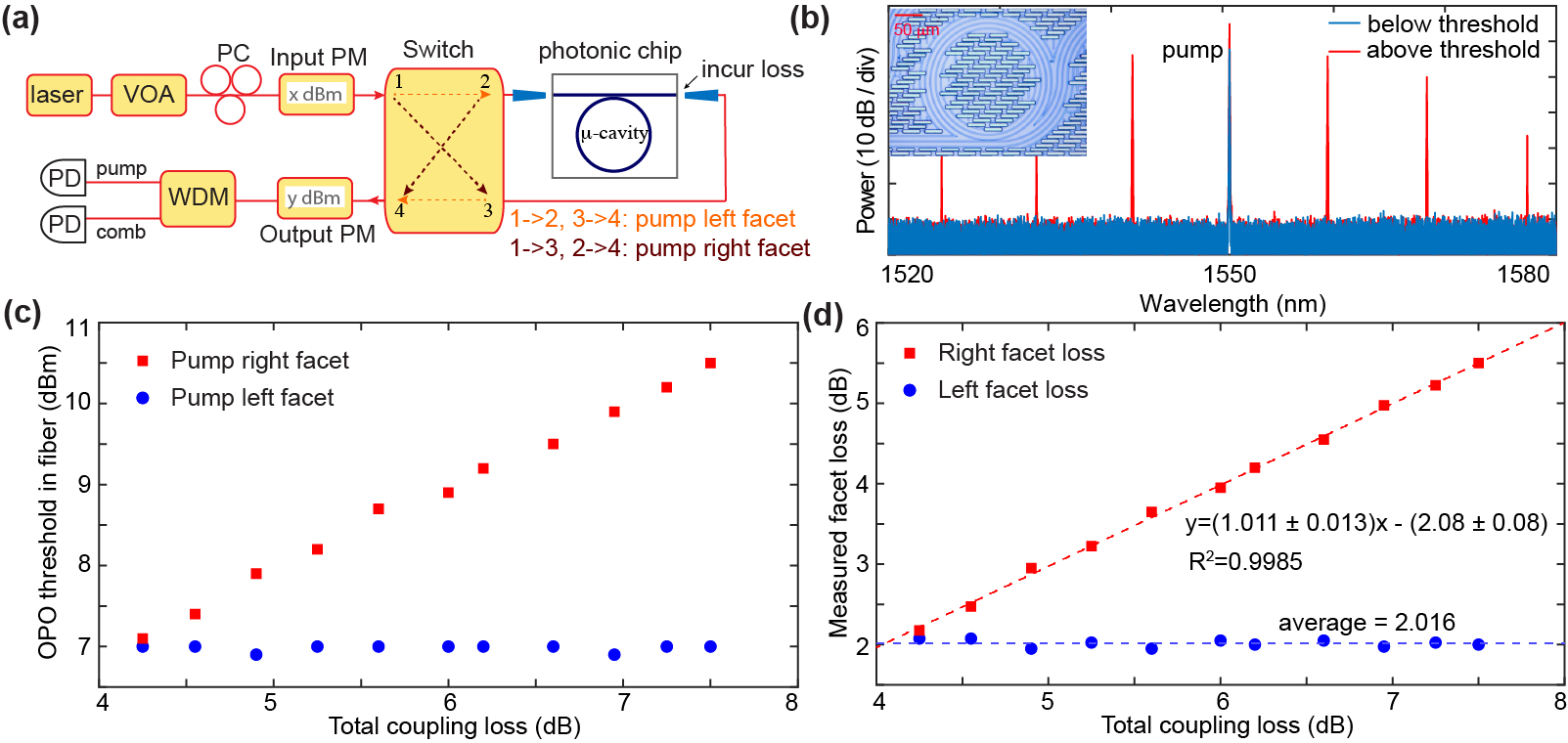}
\caption{{\bf Validation of the loss and gain measurement method.} 
    \textbf{(a)} Simplified setup for facet loss measurement. The photonic chip is pumped by an amplified single-wavelength laser. An optical switch is used to route the laser to pump the left facet or the right facet. To measure the OPO comb threshold conveniently, a wavelength division multiplexer (WDM) is used to separate the pump wavelength from the comb spectrum. The right coupling facet can be intentionally misaligned to incur additional loss. \textbf{(b)} Optical spectrum of the microresonator output when the pump laser power is below (blue) and above (red) the OPO threshold. Inset: microscopic image of the OPO microresonator. \textbf{(c)} OPO threshold measurements when additional loss is deliberately incurred at the right coupling facet. The OPO threshold when pumping the left facet (blue) stays unchanged, while the OPO threshold when pumping the right facet (red) increases with the incurred loss.
    \textbf{(d)} The measured left facet loss (blue) and right facet loss (red) versus the total coupling loss when additional loss is incurred at the right facet. The left facet loss stays unchanged, while the right facet loss increases linearly with the incurred loss, with a slope of 1. This validates that our method can clearly identify that the incurred loss is coming from the right facet.}
\label{fig:facet}
\end{figure*}

Fig.\ref{fig:concept}a illustrates the measurement method of individual coupling facet. Previously, even with the help of OBR \cite{bauters2011planar}, only the total loss of two coupling facets (fiber to fiber loss) can be measured in a photonic circuit, without knowing the loss of each coupling facet\cite{bauters2011ultra,bauters2011planar,lee2012ultra}.
In our method, a nonlinear microcavity is placed between two coupling facets. The microcavity can act as an optical parametric oscillator \cite{kippenberg2004kerr}, and its intrinsic OPO threshold does not depend on the direction of the pump laser. We then measure the in-fiber OPO thresholds pumping from the left facet ($P_L$ dBm) and pumping from the right facet ($P_R$ dBm). The in-fiber threshold power is read out from an inline power meter placed before the coupling facet (Fig.\ref{fig:facet}a). If the in-fiber OPO threshold pumping from the left facet is $\Delta$ dB higher than that of the right facet ($P_L - P_R = \Delta$ dB), then we know the left facet has $\Delta$ dB more loss than the right facet ($\alpha_L-\alpha_R = \Delta$ dB). With a loss measurement of total fiber-to-fiber optical path ($\alpha = \alpha_L+\alpha_R$ dB), we can arrive at the left and right facet loss of
\begin{equation}
    \alpha_L = \frac{\alpha + \Delta}{2}, \alpha_R = \frac{\alpha - \Delta}{2}
    \label{eq:facet}
\end{equation}
For simplicity, all power and loss in this paper are expressed in log scale, as dBm or dB. The propagation loss of the short, low-loss waveguide is neglected here.

The facet loss measurement can be extended to a universal case: measurement of unknown loss ($\alpha_X$) for a device under test (DUT) on a chip (Fig. \ref{fig:concept}b). A pair of nonlinear microcavities (labeled 1 and 2) can be placed on the left and right of the DUT, and their in-fiber OPO thresholds are measured when the laser is pumping at the left facet ($P_{L,1}$, $P_{L,2}$) and right facet ($P_{R,1}$, $P_{R,2}$). Similar to the facet loss measurements above, in-fiber OPO threshold difference on microcavity 1 can give us the loss difference on the left and right of cavity 1: $\Delta_1 = P_{L,1} - P_{R,1} = \alpha_L - (\alpha_X + \alpha_R)$. And for microcavity 2, we can have a similar relationship: $\Delta_2 = P_{L,2} - P_{R,2} = \alpha_L + \alpha_X - \alpha_R$. Fiber to fiber total optical path loss, $\alpha =\alpha_L + \alpha_X + \alpha_R$, can also be directly measured. As a result, loss for DUT, left facet and right facet can all be extracted: 
\begin{equation}
    \alpha_X = \frac{\Delta_2 - \Delta_1}{2}, \alpha_L = \frac{\alpha + \Delta_1}{2}, \alpha_R = \frac{\alpha - \Delta_2}{2}.
    \label{eq:DUT}
\end{equation}
In this way, all losses in the circuit can be identified without disassembling the photonic circuits. An example is given in Fig.\ref{fig:concept}c, where the loss of the left facet, right facet, and a device under test (compact waveguide crossing) are identified. The measured loss for the compact waveguide crossing is 1.35 dB, which is within the range of $1.3 \pm 0.05$ dB given in the Foundry library.

The simplified experiment schematic is shown in Fig. \ref{fig:facet}a. An amplified continuous-wave (cw) laser is used to probe the photonic chip. A voltage-controlled optical attenuator (VOA) and an in-line optical power meter (PM) are used to control and monitor the pump power, respectively. To conveniently pump the photonic chip from the left facet as well as the right facet, a four-port electrically-controlled optical switch is used to route the pump laser. When the switch is set to port 1 to port 2, and port 3 to port 4, the pump laser enters the chip from the left facet and exits from the right facet. When the switch is set to port 1 to port 3, and port 2 to port 4, the pump laser enters the chip from the right facet and exits from the left facet. A second in-line power meter is placed after the optical switch to measure the fiber-to-fiber total optical path loss. The loss of the optical switch can be pre-calibrated and subtracted from the loss measurement. 

The nonlinear process used here for loss measurement is microcavity optical parametric oscillation, which is based on degenerate four-wave mixing induced by optical Kerr nonlinearity \cite{kippenberg2004kerr}. Microcavity OPO has been widely studied in the past two decades for microresonator-based frequency comb generation\cite{del2007optical,kippenberg2011microresonator,herr2014temporal, gaeta2019photonic}. In the OPO process, when the pump laser exceeds the threshold power, pairs of OPO sidebands will be generated at microcavity resonance frequencies that are multiple free-spectral-range (FSR) away from the pump laser \cite{herr2012universal} (Fig. \ref{fig:facet}b). Because of the existence of a well-defined threshold power, microcavity OPO can serve as a great optical power discriminator for the loss measurement. 



\noindent {\bf Method validation and method precision:} To validate the loss measurement method, the measured loss has to be compared to a known loss. However, this is not straightforward, as no other method exists for component-level loss measurement. Here, we create a scenario where this comparison is possible. In the facet loss measurement, we can incur additional loss on a facet, e.g., right facet, by misaligning the lensed fiber to the waveguide while leaving the rest of the optical path unchanged. The additional loss can be directly measured by monitoring the total loss of the optical path. If our method is accurate, the loss of the right facet measured by our method should increase by the same amount as the total loss, while the measured loss of the left facet stays the same. 

\begin{figure*}[!bht]
\captionsetup{singlelinecheck=off, justification = RaggedRight}
\includegraphics[width=17cm]{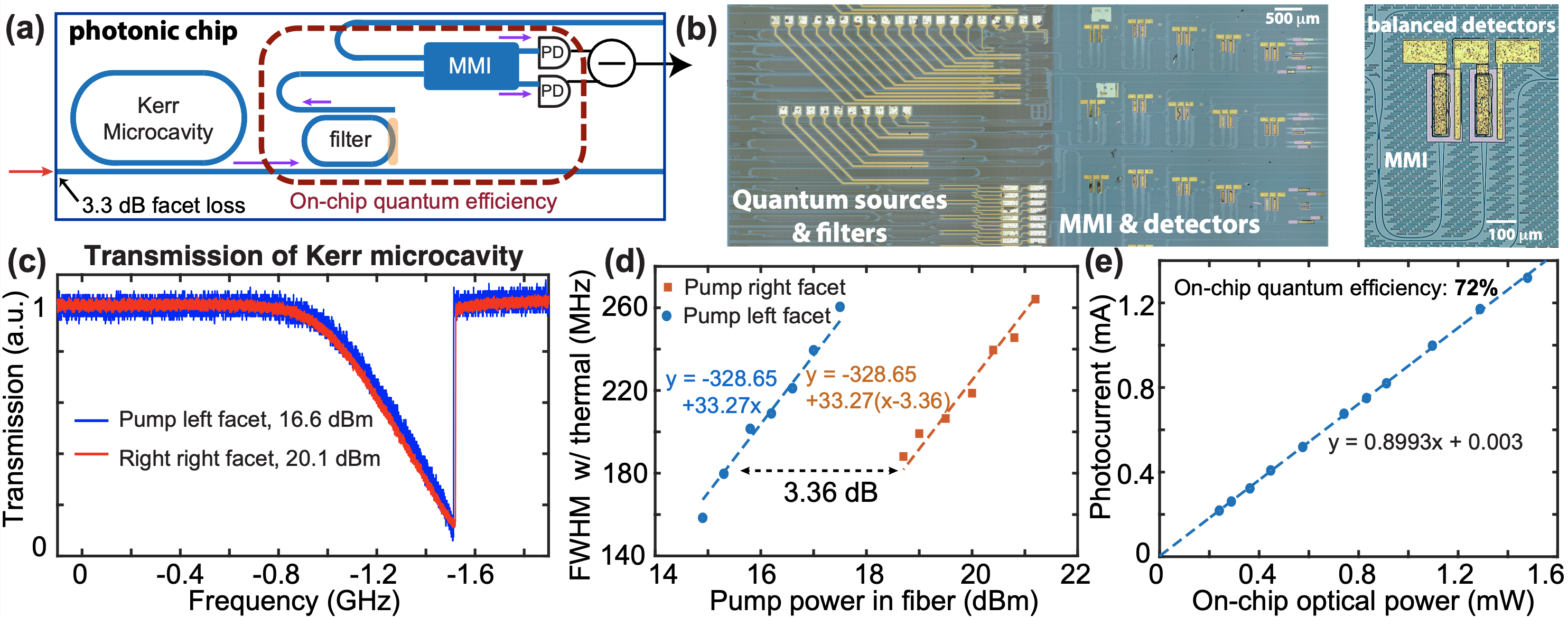}
\caption{{\bf Measurement of on-chip efficiency of a quantum photonic circuit using thermal-optics nonlinearity.} 
    \textbf{(a)} Circuit illustration of the quantum photonic chip. A Kerr microcavity serves as a broadband entanglement and squeezing source, followed by a ring filter to pick up quantum modes at selective wavelengths. The drop-port of the filter is then combined with a local oscillator arm on a 50/50 MMI coupler, and the MMI output goes into a pair of balanced photodiodes heterogeneously integrated on the photonic chip. The efficiency after the Kerr microcavity (the dashed box) directly impacts the entanglement and squeezing quality. This efficiency can be conveniently extracted after the loss measurement of the left coupling facet.  
    \textbf{(b)} Microscopic pictures of the heterogeneously integrated photonic chip, and a zoom-in image of a typical pair of integrated balanced photodiodes. \textbf{(c)} Transmission of the Kerr microcavity when pumping from the left facet (blue, 16.6 dBm in fiber power) and right facet (red, 20.1 dBm in fiber power). The triangle-shaped resonance transmission is caused by the thermal-optics effect of the pump laser. \textbf{(d)} In-fiber pump power versus thermal broadened full-width at half maximum (FWHM) of the transmission resonance. On average, the pump power at the right facet needs to be 3.36 dB higher than the pump power at the left facet to reach the same thermal broadening of the resonance transmission. \textbf{(e)} Photocurrent on the balanced photodiodes versus the on-chip optical power. On-chip optical power is obtained by correcting the in-fiber power with the 3.3 dB left facet loss.}
\label{fig:QE}
\end{figure*}

We implement this comparison in our experiment, and the setup is shown in (Fig. \ref{fig:facet}a). When the incurred loss increases on the right facet, the measured in-fiber OPO threshold increases for the right facet pumping but remains the same for the left facet pumping (Fig. \ref{fig:facet}c). The facet loss can then be obtained using eq. \ref{eq:facet} and is presented in Fig. \ref{fig:facet}d. The measured right facet loss increases linearly with the measured total loss, with a slope of $1.01 \pm 0.01$, while the measured left facet loss stays unchanged. This validates that our method can indeed identify the location of the loss in the optical path with very good accuracy. The precision of the facet loss measurement method can also be inferred from this measurement. The same left facet loss was measured 11 times with different right facet loss, and we get an average 2.016 dB left facet loss, with a standard deviation of 0.045 dB. This can represent the precision of the method.

\noindent {\bf Method accuracy:} Optical loss/gain measurements relies on the measurement of optical power. While commercial optical power meters can have high precision and sensitivity (small standard deviation), most of them have an accuracy uncertainty of $\pm 5\%$, meaning that the measured optical power can be $\pm 0.2$ dB off from the true value. It is thus important to inspect whether our loss/gain measurement method is immune to the accuracy uncertainty of the optical power meters. In our method, the DUT loss $\alpha_X=(\Delta_2 - \Delta_1)/2$ is measured from the difference of in-fiber OPO thresholds. In our experimental setup, all OPO thresholds are read out from the same input power meter before the optical switch. As a result, the absolute offset of the power meter is canceled in $\alpha_X$ and will not affect the accuracy. 
For the facet loss, the fiber-to-fiber loss of the entire optical path has to be measured, which involves two power meters. The impact of the power meters' absolute offset difference can be eliminated by an additional protocol:  do the measurement twice and swap the input power meter and output power meter. The average of the two measurements will cancel out the offset difference in the two power meters.

\vspace{12pt}

\noindent {\bf On-chip quantum efficiency measurement.}
The method demonstrated above is then applied to measure the true quantum efficiency for on-chip photodiodes and photonic circuits. In quantum optics, measurements play a central role in almost all tasks, ranging from quantum teleportation \cite{pirandola2015advances} to measurement-based quantum computing\cite{briegel2009measurement,o2007optical}.
The efficiency of the optical path after the quantum source directly impacts the quality of squeezing and entanglement. It is thus critical to know the overall on-chip quantum efficiency of the circuit, including the transmission of filters, couplers, and the quantum efficiency of photodiodes. 

Here we demonstrate such a measurement for the first time, to our knowledge. In our photonic circuit, a Kerr microcavity can serve as a quantum emitter for photon pairs or squeezed light\cite{dutt2015chip,reimer2016generation,vaidya2020broadband,zhao2020near,yang2021squeezed,jahanbozorgi2023generation}. Then, a drop filter is followed to route the photons to be combined with a local oscillator on a 50/50 MMI beam splitter, and detected on a pair of heterogeneously integrated photodiodes \cite{yu2020heterogeneous,gao2025heterogeneous}  (Fig. \ref{fig:QE}a,b).
To measure the on-chip quantum efficiency after the quantum emitter, we first obtain the external quantum efficiency by measuring the in-fiber optical power (off-chip power) versus photocurrent of the photodiodes. Then we use the nonlinearity of the Kerr microcavity to measure the left facet loss. We then can obtain the on-chip optical power, which gives us the on-chip quantum efficiency of the circuits. In this circuit, we measured the fiber to fiber loss of 9.9 dB, left facet loss of 3.3 dB, and an on-chip quantum efficiency of $72\%$ (Fig. \ref{fig:QE}e). The optical power meter has a $\pm 5\%$ uncertainty for the absolute optical power measurement, which will be passed onto the quantum efficiency. It should be noted that without this method, researchers often assume equal facet loss on the left and right, and this will give an incorrect left facet loss of $4.95$ dB, which then overestimates the on-chip quantum efficiency to 105$\%$. 

Finally, we show that our method can be easily implemented with the thermal-optics effect\cite{carmon2004dynamical}, which is widely observed on many photonic platforms. When the pump power is coupled into the microcavity, the microcavity is heated up by the light absorption inside the microcavity, and the rise of temperature leads to the change of the refraction index, which shifts the resonance frequency. This nonlinear effect leads to a thermal triangle in the resonator transmission, which can be easily observed by scanning the laser frequency across the resonance frequency \cite{carmon2004dynamical} (Fig. \ref{fig:QE}c). The width of the thermal triangle increases monotonically with the input optical power and thus can be used as an on-chip power discriminator. 

The left facet loss of the quantum circuit chip (Fig. \ref{fig:QE}b) is measured using the thermal-optics effect. The resonance transmission of the Kerr microcavity is shown in Fig. \ref{fig:QE}c. When reaching a similar thermal triangle, the in-fiber power when pumping the left facet is 16.6 dBm, while the in-fiber power when pumping the right facet is 20.1 dBm. This suggests the right coupling facet has roughly 3.5 dB more loss than the left coupling facet. The thermal triangle width (full width half maximum linewidth) can be measured at a series of pump powers, and an average of 3.36 dB facet loss difference can be obtained (Fig. \ref{fig:QE}d). When combined with total facet loss of 9.9 dB, we obtained 3.3 dB facet loss on the left.

\medskip
\noindent {\bf Discussion.}

We make a brief comparison of our method and the OBR method \cite{barfuss1989modified,soller2005high}, which are fundamentally distinct. The principle of OBR is similar to that of the coherent laser sweeping LiDAR, where a fast frequency sweeping laser is launched to the object and its reflection is coherently detected to determine the distance and reflection strength. As back-relection is not the only source of optical loss, OBR can only infer optical loss indirectly by examining the back-reflection strength before and after the DUT. This requires the waveguide or fiber to be identical before and after the DUT, and also sufficient waveguide length to perform statistic average of the weak reflection signal. This excludes coupling loss measurements of individual facet, as the waveguide/fiber is different before and after the coupling facet. In addition, OBR's working wavelength is limited to the wavelengths where high quality fast sweeping lasers exist. Also, the use of broadband frequency sweeping laser will cause difficulty for inspecting circuits with narrow-bandpass filters. The loss measurement method described in this manuscript does not have those limitations. Our method shows an uncertainty better than 0.1 dB, which is significantly better than the indirect loss characterization using OBR so far\cite{bauters2011ultra,bauters2011planar,lee2012ultra}.  While additional ring resonators are needed in the photonic circuits in our method, they are very mature in today's photonic platforms, including Silicon photonics, silicon nitride and thin-film lithium niobate\cite{zhang2017monolithic}. Its impact on the photonic circuit design and footprint can be minimized, as ring resonators can be made sufficiently small, e.g., radius of tens of micrometer.

In summary, we have demonstrated a general method to characterize loss and gain inside the photonic integrated circuits. Loss of individual facet, unknown devices under test, and the quantum efficiency of a heterogeneously integrated circuit are measured. Although only loss measurement is shown in this manuscript, on-chip gain can be characterized by the exact same protocol. The method can be immediately applied to characterize loss and gain in a variety of scenarios in integrated photonics to understand photodetection quantum efficiency, quantum emitter efficiency, and nonlinear frequency conversion efficiency, e.g., second-harmonic generation in thin-film lithium niobate\cite{wang2018ultrahigh,lu2019periodically}. In the long term, we anticipate that the method will improve our understanding of loss and gain of integrated photonic components, and ultimately improve the design and implementation of large-scale photonic integrated circuits and their applications.

\medskip

\noindent\textbf{Methods}
\begin{footnotesize}

\noindent{\bf OPO threshold measurement.} 
To conveniently measure the OPO threshold in the experiment, a fiber Bragg grating (FBG) filter is used as a wavelength division multiplexer to separate the light at the pump laser wavelength and other wavelengths. Photodiodes (PDs) are used to measure the pump laser transmission and detect the comb generation (OPO generation). In the threshold measurement, laser frequency is ramped around the microcavity resonance frequency, and the transmission and comb power are monitored on the oscilloscope. The pump power is increased until the comb power is detected on the photodiode. In-fiber OPO thresholds can be measured conveniently using this protocol.

\end{footnotesize}

\medskip

\noindent\textbf{Acknowledgement}

\noindent The authors acknowledge Ligentec for SiN microresonator fabrication, and gratefully acknowledge DARPA INSPIRED (HR0011-24-2-0360), DARPA GRYPHON (HR0011-22-2-0008), DARPA NaPSAC (N660012424000), National Science Foundation (1842641, 2238096), DOE (DE-SC0023337) and QC82 Inc. The views and conclusions contained in this document are those of the authors and should not be interpreted as representing official policies of DARPA, DOE, or the U.S. Government.

\medskip

\noindent \textbf{Author Contributions}\\ 
X.Y. conceived the concept of the experiments. H.C., R.L., G.K. performed the measurements. F.T, X.G., A.B. designed and fabricated the integrated photodiodes. G.K., R.L., H.C., Z.Y., S.S., B.W. designed the photonic integrated circuits. H.C., R.L., G.K., X.Y. analyzed the experimental results. X.Y. supervised the experiments. All authors participated in preparing the manuscript.

\medskip

{\noindent \bf Competing interests}
The authors declare no competing interests.

\medskip

{\noindent \bf Data availability.} The data that support the plots within this paper and other findings of this study are available from the corresponding author upon reasonable request. Accession to all relevant data will be available online before publication.


\medskip

{\noindent \bf Code availability.} The codes that support the findings of this study are available from the corresponding authors upon reasonable request.

\bibliographystyle{naturemag}
\bibliography{ref}

\end{document}